\documentclass[conference]{IEEEtran}
\usepackage{algorithm}
\usepackage{algorithmic}
\usepackage{url}
\usepackage{amsthm}
\usepackage{amssymb}
\usepackage{amsmath}
\usepackage{epsfig}
\usepackage{epstopdf}
\usepackage{subfigure}
\usepackage{balance}
\usepackage{multirow}

\usepackage{color}
\usepackage[table]{xcolor}
\usepackage{color, colortbl}

\newtheorem{prop}{Proposition}

\definecolor{Gray}{gray}{0.9}
\definecolor{LightCyan}{rgb}{0.88,1,1}

\ifCLASSINFOpdf
\else
\fi
\begin{document}
%
\title{Resource Provisioning and Profit Maximization for Transcoding in Information Centric Networking }

\author{\IEEEauthorblockN{Guanyu Gao, Yonggang Wen}
\IEEEauthorblockA{School of Computer Engineering\\
Nanyang Technological University\\
\{ggao001, ygwen\}@ntu.edu.sg}
\and
\IEEEauthorblockN{Cedric Westphal}
\IEEEauthorblockA{Huawei Innovation Center \&\\
University of California, Santa Cruz\\
cedric.westphal@huawei.com, cedric@soe.ucsc.edu}}


%



\maketitle

\begin{abstract}
Adaptive bitrate streaming (ABR) has been widely adopted to support video streaming services over
heterogeneous devices and varying network conditions.
With ABR, each video content is transcoded into multiple representations in different bitrates and resolutions.
However, video transcoding is computing intensive, which requires the transcoding service providers to deploy a large number of servers for transcoding the video contents published by the content producers.
As such, a natural question for the transcoding service provider is how to provision the computing resource for transcoding the video contents while maximizing service profit.
To address this problem,
we design a cloud video transcoding system by taking the advantage of cloud computing technology to elastically allocate computing resource.
We propose a method for jointly considering the task scheduling and resource provisioning problem in two timescales, and formulate the service profit maximization as a two-timescale stochastic optimization problem.
We derive some approximate policies for the task scheduling and resource provisioning.
Based on our proposed methods,
we implement our open source cloud video transcoding system \emph{Morph}
and evaluate its performance in a real environment. The experiment results demonstrate that our proposed method can reduce the resource consumption and achieve a higher profit compared with the baseline schemes.
\end{abstract}



%
\IEEEpeerreviewmaketitle

\section{Introduction}

With the emergence of Information-Centric Networking (ICN), the future Internet could shift away from a point-to-point paradigm to a more content-centric one \cite{fayazbakhsh2013less}.
In this condition, the content producer can focus on providing the information object of the video content.
Nowadays, adaptive bitrate streaming has been widely adopted to deliver the video content in different bitrates and resolutions to adopt varying network conditions and heterogeneous devices \cite{wen2014cloud}.
Video transcoding is used to pre-transcode the original contents into multiple representations.
For each video content published by the content producers, the transcoding service provider will transcode it into multiple representations in different bitrates and resolutions,
which will be further stored in the original servers and the other caching nodes in the network. As such, the viewers can request the multiple representations of the video contents in ABR.
Some transcoding function placed in the network can decide and dynamically adjust the optimal number of representations for the video contents and the places for caching to be accessed in high quality.
With the named function networking (NFN) framework, the video content can also be transcoded on the fly \cite{tschudin2013named}.
%

Implementing ABR yields a cost for both transcoding service provider and content producer.
First, video transcoding is computing intensive, consuming a huge amount of computing resource.
The transcoding service providers need to purchase and maintain a large number of servers for transcoding to meet the peak workload. However, the servers can only be set in the idle state when no transcoding tasks are performed, wasting too much computing resource.
Second, video transcoding takes excessive time and the content producers need to wait for a long time for content publishing, which is intolerable for the contents needing timely delivery.
The emergence of cloud computing technology introduces an opportunity to improve video transcoding \cite{wen2014cloud}.
By leveraging the cloud infrastructure, it can transcode video contents more efficiently compared with the traditional  transcoding methods. Specifically, it can perform multiple transcoding in parallel using a large number of virtual machine (VM) instances or containers, which can greatly reduce the transcoding time. Meanwhile,
it can elastically provision the computing resource according to the transcoding workload, which can avoid the resource wastage and reduce monetary costs.


The current research on the video transcoding system mainly focuses on how to reduce the energy consumption and processing delays \cite{Song:2015, ma2014dynamic, timmerercloud}, without much attention on the resource provisioning problem in the cloud.
In this paper, we consider how to provision the computing resource for transcoding video contents while maximizing the profit for the transcoding service provider by taking the advantage of the cloud computing paradigm.
%
To this end, we mainly cope with the following system management problems in video transcoding system.
First, as the system workload is time-varying, we assess how to dynamically provision the resource for the transcoding service.
Second, with a certain amount of provisioned computing resource and a number of transcoding tasks to be processed, we consider how to schedule the tasks to maximize the revenue while meeting the service requirements.
Moreover, the complexity of the video data makes it hard to precisely estimate the transcoding time for a task, which is essential for the task scheduling and resource provisioning.

To solve the above problems, we first propose a neural network method for precisely estimating the transcoding time of the tasks.
We then propose a method for jointly considering the task scheduling and resource provisioning problem for the system management, and formulate it as a two-timescale Markov Decision Process \cite{chang2003multitime}. 
We derive the policies for task scheduling and resource provisioning for maximizing the overall profit.
Based on our design, we implement the system and present the code of the practical implementation in Github \cite{akilos}.
We evaluate the system performance in a real environment.

\section{System Architecture and Workflows} \label{sec_3}
\begin{figure}
\begin{center}
\epsfig{file=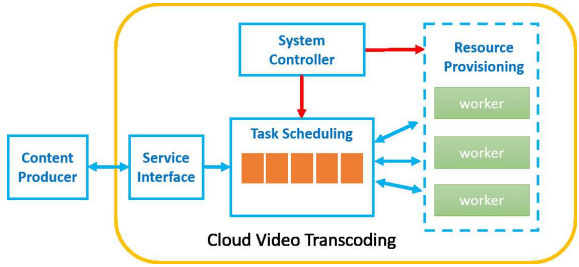, width=0.85\columnwidth}
\end{center}
\caption{The system architecture of the cloud video transcoding system.}\label{fig:sys-arch}
\end{figure}

\subsection{System Architecture}
We illustrate the system architecture in Fig. \ref{fig:sys-arch}.
The functionalities of each module are detailed as follows:

\emph{Service Interface Module:}
It serves as the interface for processing the transcoding requests from the content producers.
The content producer uploads the video content and submits the transcoding request as a transcoding task.
For each transcoding task, the system will estimate the required computing resource and divides the original video content into independent video blocks according to the GOP structure.

\emph{Resource Provisioning Module:}
It provisions a number of homogenous VM instances in the transcoding cluster\footnote{It can also adopt the Container technology for resource virtualization.}. Each VM instance runs a transcoding worker.
The resource provisioning module dynamically adjusts the number of active VM instances in the transcoding cluster according to the system workload and the resource provisioning policy.

\emph{Task Scheduling Module:}
It maintains a queue for the pending transcoding tasks and adjusts their transcoding order by reordering the tasks
according to the task scheduling policy.
When there is a request for a video block from the transcoding worker, the task scheduler will pick a video block of the head-of-queue task to dispatch.
Once a transcoding task has been started to perform, i.e.,
the first video block of the transcoding task has been dispatched for transcoding, the task will stay at the head-of-queue until all of its video blocks have been dispatched to the transcoding workers.
Then, the next request for a video block from the transcoding worker will be replied with the video block of the next head-of-queue task.
%
%

\subsection{System Workflows}
The workflows for fulfilling the transcoding requests from the content producers are as follows.
\emph{First,} the content producer uploads a video file and submits the transcoding task by specifying the service requirements. 
The system will estimate the required transcoding time for the new task and split the video file into video blocks. Each of the video block consumes the same amount of computing time.
\emph{Second,}
for a number of pending transcoding tasks in the queue, the task scheduler reorders the tasks periodically according to the task scheduling policy.
\emph{Third,}
when a transcoding worker becomes idle, it will request a video block from the task scheduler.
After transcoding the video block into the target representation,
the transcoding worker will send the transcoded video block back to the task scheduler. The task scheduler will concentrate the transcoded video blocks into one video file after receiving all of the transcoded video blocks of a transcoding task from the transcoding workers.
The transcoded video contents can be stored in the content network for delivery.

\section{System Model and Problem Formulation} \label{sec_4}

\subsection{Task Arrival Model}

We adopt a discrete time model and divide the time horizon into two timescales. We denote the time in the fast timescale as $t \ (t=0,1,2,...)$, and the time in the slow timescale as $N_{k} \ (k =0,1,2,...)$.
Each $N_k$ consists of $T$ time slots in the fast timescale, and
the time duration of $N_{k}$ in the slow timescale is from the time slot $kT$ to the time slot $(k+1)T-1$. The typical length of one time slot is $1\sim10$ seconds and $T$ is $1800 \sim 3600$ seconds.
We model the task arrival as a non-stationary Poisson process with different arrival rates across the slow timescales. For each time slot within $N_k$, we assume the task arrival distribution is homogenous.
We denote the task arrival rate in each time slot $t$ of $N_k$ as $\lambda_k$, $\ kT \leq t < (k+1)T$.

\subsection{Transcoding Time Estimation Model}
For each of the transcoding tasks,
we need to estimate the required computing resource (i.e., transcoding time of a task) for the task scheduling.
We adopt the neural network method for learning the non-linear relationship among the transcoding time of a task and other dependent factors.
We use a 3-layer feedforward neural network, which consists of one input layer, one hidden layer, and  one output layer.
%
We denote the input feature vector of the input layer as $\xi$.
To construct $\xi_i$ for the transcoding task $i$, we use the duration, bitrate, frame rate, and resolution of the original video file, and the resolution of the target video file as inputs. The output layer generates the estimated transcoding time for this task. The nonlinear relationship between the estimated transcoding time and the input feature vector is
\begin{equation}
D_i = \Gamma(\xi_i),\\
\end{equation}
where $\Gamma$ is the neural network trained offline before running the system and $D_i$ is the estimated transcoding time of task $i$.

\subsection{Service Revenue Model}
%
%
We adopt a pricing mechanism which involves the task consumed computing time, task completion time, and the service level.
The valuation function of a transcoding task is given in Eq. \ref{valuation_function}.
For the transcoding task $i$, which arrives at $a_i$, the revenue gained from this task if completed at time  $t$ is
\begin{equation}
U_i(t) = \alpha^{t-a_i} R_i D_i, 0 < \alpha < 1, t \ge a_i, \label{valuation_function}
\end{equation}
where $\alpha$ is the price discounting factor, $R_i$ is the initial marginal price for one unit of computing time required for transcoding, and $D_i$ represents the amount of computing resource to be provided by the service provider.
%
%
The form of the valuation function will affect the derivation of the task scheduling policy in the fast timescale.
Our method can also be applicable to some other functions, e.g., linear functions and step functions. The valuation function of linear form is:
\begin{equation}
{U'}_i(t) = w_i - \beta_i (t-a_i), t \ge a_i, \beta_i > 0, \tag{\theequation.a} \label{linear_decreasing_function}
\end{equation}
where $w_i$ is initial revenue for task $i$ and $\beta_i$ is the discounting factor. The values for ${U'}_i(t)$ which are less than zero can be seen as service penalty for the processing delay. The step valuation function is:
\begin{equation}
{U''}_i(t) = \begin{cases}
w_i, & a_i \le t \le a_i + \tau_i, \tag{\theequation.b} \label{step_decreasing_function} \\
0,       & t \ge a_i + \tau_i,
 \end{cases}
\end{equation}
where $\tau_i$ is the service deadline for task $i$. If the task miss the deadline, the revenue will be zero.

%
%
\subsection{VM Instance Cost Model}
The system needs to dynamically adjust the number of provisioned VM instances for resource provisioning.
Nevertheless, provisioning a new VM instance consumes substantial time, thus the operation should not be done too frequently.
As such, we scale the transcoding cluster each $T$ time slots 
at the beginning of each $N_k$, $k=1,2, ...$, and also $T$ is not too large to ensure that the task arrival rate is relatively constant over the $T$ time slots.
%
We assume that the VM instances in the transcoding cluster are homogeneous and the overall cost is proportional to the number of provisioned VM instances.
We denote the number of provisioned VM instances in the transcoding cluster during $N_k$ as $M(N_k)$. As such,
the total cost for provisioning the VM instances at $N_k$ is
\begin{equation}
C^v(N_k) = M(N_k)C_v,\\
\end{equation}
where $C_v$ is the cost for one VM instance over $T$ time slots.

%

\subsection{Transcoding Service Profit Maximization Problem}
Our objective is to maximize the overall profit of the transcoding service. The dynamics of our system can be characterized in the separable slow and fast timescale. The resource provisioning operation, acting at the lower frequency, has a relative long-term effect on the system performance. The task scheduling operation takes the resource provisioning operation as input and tracks the revenue maximization of the tasks at the higher frequency. We jointly consider the task scheduling and resource provisioning in two timescales.
%
%
%
%
We define the system state at the beginning of each $N_k$ in the slow timescale as
$\psi^s_{k} = \{\lambda_k, m_{k-1}\}$, where $\lambda_k$ is the discrete task arrival rate for each time slot of $N_k$
and $m_{k-1}$ is the number of active VM instances before taking the resource provisioning operation.
The state space for the system state in the slow timescale is denoted as $\Psi^s$.
We define the system state in the fast timescale at the time slot $t$ as $\psi^f_{t} =  \{x_t\} $, where $x_t$ is the set of pending tasks at time slot $t$.
For each pending task, it has the information of the elapsed time from its submission and its estimated required computing time.
The state space for the system state in the fast timescale is denoted as $\Psi^f$.
We assume that the system state space in the slow timescale and the fast timescale are both finite.

The resource management policy determines the number of VM instances to be shut down or activated at the beginning of each $N_k$ according to the system state in the slow and fast timescale.
We denote the resource provisioning policy in the slow timescale as $\pi^s$ and
the resource provisioning operation at $N_k$ as $\nu_k$. Specifically, $\nu_k > 0$ represents the number of new activated VM instances and $\nu_k < 0$ represents the number of shutdown VM instances.
The finite action space for the resource provisioning operation is denoted as $\wedge$.
The mapping from the system state in the slow and fast timescale to the resource provisioning operation by applying the policy $\pi^s$ is
\begin{equation}
\pi^s: (\psi^s_{k}, \psi^f_{kT}) \to \nu_k, k = 0,1,2,...
\end{equation}
The number of active VM instances after taking the resource provisioning operation is $m_k$, where $m_k = m_{k-1} + \nu_k \ge 0$.

In our system, the task scheduling policy in the fast timescale  determines the transcoding order of the pending transcoding tasks to maximize the overall revenue.
%
%
%
The system state in the fast timescale evolves over $T$ time slots until the system state in the slow timescale changes.
The system dynamic in the fast timescale is an MDP over finite $T$-horizon and the task scheduling policy is a sequence of $T$-horizon nonstationary policies. We define the sequence of task scheduling policies over the finite $T$-horizon as
$\pi^f_T$, and $\pi^f_{T}(t)$ is the task scheduling policy at the time slot $t$.
At each time slot $t$, we define the mapping from the system state and operation in the slow timescale and the system state in the fast timescale to the task scheduling operation for the pending tasks as
\begin{equation}
\pi^f_T(t): (\psi^s_{k}, \nu_k, \psi^f_{t}) \to \ell_t, \ kT \leq t < (k+1)T,
\end{equation}
where $\ell_t$ is the task scheduling operation for the pending tasks.
Specifically, in our system, $\ell_t$ is the scheduled transcoding order of the pending transcoding tasks in the queue.

At the beginning of $N_k$,
given the slow timescale state $\psi^s_{k}$, the resource provisioning operation $\nu_k$, the fast timescale state $\psi^f_{kT}$, and the $T$-horizon task scheduling policy $\pi^f_T$,
the total expected service profit over $T$ time slots in $N_k$ is
\begin{align}
R^s_k(\psi^s_{k}, \psi^f_{kT}, \nu_k, \pi^f_T) & = 
 \mathbb{E}_{\psi^f_{t}} \Bigg\{ \sum_{t=kT}^{(k+1)T-1} P(t) - C^v(N_k)  \Bigg\}, \nonumber
\end{align}
where $P(t)$ is service revenue at the time slot $t$.
%
%
We aim to maximize the overall future discounted profit by applying appropriate task scheduling policy $\pi^f_T$ and resource provisioning policy $\pi^s$.
Mathematically, we present the Service Profit Maximization (SPM) Problem as follow
\begin{align}
\mathcal{P}1:\  \  \max_{\pi^s \in \Pi^s} \max_{\pi^f_T \in \Pi^f_T} \mathbb{E}_{\psi^s_{k}, \psi^f_{t}} \left\{ \sum_{k=0}^{\infty} \gamma^k R^s_k (\psi^s_{k}, \psi^f_{kT}, \nu_k, \pi^f_T) \right\}, \nonumber
\end{align}
where $\gamma$ is the discounting factor, and $\Pi^s$ and $\Pi^f_T$ are the finite set of possible resource provisioning policies and task scheduling policies, respectively.
To derive the optimal policies for maximizing the overall profit,
one can in principle derive the offline policies with the method of value iteration.
In a practical system, however, the system state space is large and the state transition probability is difficult to obtain exactly. We will discuss the approximate methods to derive the task scheduling policy and resource provisioning policy.

\section{Approximate Policies for Service Profit Maximization Problem} \label{sec:approximate_solution}
In this section, we present some approximate policies for task scheduling and resource provisioning.

\subsection{Value-based Task Scheduling Policy}

Our approximation for the task scheduling policy is to assume that the number of active VM instances is unchanging for the current set of pending tasks, and the task scheduler determines the task transcoding order $\ell_t$ of the pending tasks by maximizing the overall revenue of the existing tasks, mathematically, we present it as
\begin{equation}
\mathcal{P}2: \max_{\ell_t} \sum_{i \in x_t} U_i(f_i),\\
\end{equation}
where $f_i$ is the completion time of task $i$ given the transcoding order $\ell_t$ of the pending tasks.
To solve this problem, we first introduce a method for estimating the completion time of the pending tasks when given a transcoding order of the pending tasks and the current number of active VM instances. Based on that, we present a method for deriving the solution of $\mathcal{P}2$.

\subsubsection{Task Completion Time}
The original video files are partitioned into video blocks to be dispatched to many VM instances for parallel transcoding and each of the video blocks consumes the same amount of computing time.
We assume that each video block consumes $F$ time slots for transcoding, and for a task $i \in x_t$, it is divided into $b_i$ video blocks.
We assume that the order of task $i$ in the queue is $o_i$.
Given the order of the pending tasks in the queue, we denote the total number of video blocks of the tasks which order is not larger than $o_i$ as $g_i$.
We have the following proposition for estimating the finish time of the $g_i$-th video block (i.e., task $i$).

\begin{prop} \label{prop_1}
Suppose that the transcoding cluster has $m_k$ active VM instances and the transcoding progresses of the current video blocks on these VM instances are unknown.
%
At the time slot $t_0$, one transcoding worker becomes idle and requests the first video block in the queue. Then, the expected completion time of the $g_i$-th video block (i.e., the task $i$) is
\begin{eqnarray}
E\{f_i\} = t_0 + \frac{F}{m_k}(g_i - 1) + F.
\end{eqnarray}
\begin{proof}
Please see Appendix \ref{append_1} for the detailed proof.
\end{proof}
\end{prop}
%

%
%
%
%
%
\subsubsection{Task Scheduling Policy}
Given the set of the pending tasks $x_{t}$, we assume task $i$, $j$, $k$ are successive in the queue and the order is denoted as $(...,i,j,k,...)$. We denote the durations between the completion time of task $i$ and $j$, task $j$ and $k$ as
\begin{eqnarray}
d_{ij} = f_j - f_i,\ d_{jk} = f_k - f_j,
\end{eqnarray}
where $f_i$, $f_j$, $f_k$ are the completion time of task $i$, $j$, $k$, respectively.
Given the order $(...,i,j,k,...)$, let the expected revenue gained from task $j$ and $k$ based on Eq. \eqref{valuation_function} be $R_{jk}$.
We exchange the order of task $j$ and $k$ while keeping the order of the other tasks unchanged, let the expected revenue gained from task $j$ and $k$ in the order of $(...,i,k,j,...)$ be $R_{kj}$.
If $R_{jk} > R_{kj}$, it can be deduced that
\begin{eqnarray}
\frac{\alpha^{d_j - a_j}  R_j D_j} {1 - \alpha^{d_j}} \ge \frac{\alpha^{d_k - a_k} R_k D_k} {1 - \alpha^{d_k}}, \label{eqn:task_order}
\end{eqnarray}
where $d_j = d_{ij} = d_{kj} = \frac{F}{m_k} b_j$ and $d_k = d_{jk} = d_{ik} = \frac{F}{m_k} b_k$.
As such, we have the following proposition for task scheduling.
\begin{prop} \label{prop_2}
If the current set of pending tasks $x_t$ are conducted in the decreasing order of the weight $P_i$, we can maximize the overall revenue gained from these pending tasks,
\begin{eqnarray}
P_i = \frac{\alpha^{d_i - a_i} R_i D_i} {1 - \alpha^{d_i}}, i \in x_t \ and \ d_i = \frac{F}{m_k} b_i, \label{eqn:task_weight}
\end{eqnarray}
\end{prop}


\begin{proof}
We assume that the current set of pending tasks has been arranged by the decreasing order of  $P_i$. The transcoding order of the task $i$ is $o_i$. If we move the task $i$ from $o_i$ to $o_i'$, it can be done by iteratively interchanging task $i$ with its neighboring task until it reaches $o_i'$. Since the tasks have been in the order of decreasing $P$, each interchanging will incur a loss on the revenue according to Eq. \eqref{eqn:task_order}. Hence, it's the optimal scheduling by the decreasing order of  $P$ for maximizing the revenue of the current set of the pending tasks.
\end{proof}

The system in the fast timescale works as follow:
in each time slot $t$,
the task scheduler calculates the value $P_i$ for each task, and sorts the tasks in decreasing order of $P_i$.
%
The value of $P_i$  does not depend on $t$, therefore, the task scheduler only needs to resort the pending tasks when new tasks come in or the number of VM instances has changed.
This method can also be applied to the valuation function Eq. \eqref{linear_decreasing_function}, and we can have $P_i = \beta_i/d_i$ for sequencing the tasks in decreasing order.
With the valuation function Eq. \eqref{step_decreasing_function}, the problem is known to be NP-hard.
In this case, however, sequencing the tasks in the decreasing order of $P_i = w_i/d_i$ is still shown to be a popular and effective approximate solution.

\subsection{Learning-based Resource Provisioning Policy}
In this subsection, we introduce the
method for deriving the resource provisioning policy.
The system dynamic in the slow timescale is an MDP with the system reward defined as the service profit over $T$ time slots if given the task scheduling policy $\pi^f_T$ in the fast timescale.
As such, we can write the system dynamic in the slow timescale as
%
\begin{align}
\mathcal{P}3: \hat{V}^* & (\psi^s_{k},  \psi^f_{kT})  =   \max_{\nu_k} \Big\{ R^s_k(\psi^s_{k}, \psi^f_{kT}, \nu_k, \pi^f_T)  \nonumber \\
&  + \gamma \mathbb{E}_{\psi^s_{k+1}, \psi^f_{(k+1)T}} \big\{ \hat{V}^*(\psi^s_{k+1}, \psi^f_{(k+1)T}) \big\}     \Big\}.
\end{align}

We leverage the \emph{Q-Learning} method, which is a model-free \emph{Reinforcement Learning} technique, to find the action-selection policy for the given MDP in the slow timescale.
The learning procedures are as follows:
at the time $N_k$, the system observes the state $(\psi^s_{k},  \psi^f_{kT})$ in the two timescales, and selects the action $\nu_k$ according to a certain resource provisioning policy $\pi^s$. After $T$ time slots, the system observes the service profit $R^s_k$ gained over the $T$ time slots in the fast timescale and the new system state $(\psi^s_{k+1}, \psi^f_{(k+1)T})$ at the time $N_{k+1}$. Then, the new estimated discounted overall future profit starting from the state $(\psi^s_{k},  \psi^f_{kT})$ by taking the action $\nu_k$ can be calculated as
\begin{align}
Q^{'}((\psi^s_{k}, \psi^f_{kT}), \nu_k) & = R^s_k(\psi^s_{k}, \psi^f_{kT}, \nu_k, \pi^f) \nonumber \\
& + \gamma \max_{\nu_{k+1}} Q((\psi^s_{k+1}, \psi^f_{(k+1)T}), \nu_{k+1}), \label{update_q_value}
\end{align}
where $Q((\psi^s_{k}, \psi^f_{kT}), \nu_k)$ is the action-value and
$\nu_{k+1}$ is the optimal action that can maximize the expected overall discounted profit starting from the state $(\psi^s_{k+1}, \psi^f_{(k+1)T})$.
As such, the action-value $Q((\psi^s_{k}, \psi^f_{kT}), \nu_k)$ can be updated based on the new estimation according to following equation,
\begin{align}
Q((\psi^s_{k}, &  \psi^f_{kT}), \nu_k) = Q((\psi^s_{k}, \psi^f_{kT}), \nu_k) \nonumber \\
& + \delta_k \{Q^{'}((\psi^s_{k}, \psi^f_{kT}), \nu_k) - Q((\psi^s_{k}, \psi^f_{kT}), \nu_k)\}, \label{eqn:q_learning}
\end{align}
where $\delta_k$ is the learning rate.
After visiting each state-value enough times, the Q-learning algorithm will converge to the optimal policy.
To reduce the dimensionality of the system space,
we adopt the feature extraction method to
obtain a compact representation of the state space, which are considered as the important characteristics of the original space.
We denote the compact state space as $\Phi$, the original state $(\psi^s_{k}, \psi^f_{kT})$ can be compacted as the state
$\phi_k = (\omega_{kT}, m_{k}, \lambda_k)$,
where $\omega_{kT}$ is the summation of the valuation of the pending task at the time $kT$, $m_{k}$ is the number of active VM instances, and $\lambda_k$ is the average task arrival rate.
We replace the original system state $(\psi^s_{k}, \psi^f_{kT})$ with $\phi_k$ in Eq. \eqref{update_q_value} and \eqref{eqn:q_learning}.
We adopt the $\varepsilon$-$greedy$ method to balance the exploring and exploiting when selecting action for learning.
%
%
%
%
Based on the above discussions, our method for learning the resource provisioning policy in the slow timescale is illustrated in Algorithm \ref{alg:resource_provisioning}.

\begin{algorithm}
\renewcommand{\algorithmicrequire}{\textbf{Input:}}
\renewcommand\algorithmicensure {\textbf{Output:} }
\caption{Learning-based Resource Provisioning Policy}
\label{alg:resource_provisioning}

\begin{algorithmic}[1]
\REQUIRE ~~\\
Initialize $Q(\phi, \nu) = C$, $\forall \phi$, $\nu$. \\
Task scheduling policy $\pi^f_T$ in the fast timescale. \\
Set the maximum number of loops, M.  \\
Set k = 0. \\
\ENSURE The optimal action-value $Q^*(\phi, \nu)$. \\

\STATE{Obtain the current system state $\phi_k$ at the beginning of $N_k$}.
\REPEAT
\STATE{Select $\nu_k$ based on $\phi_k$ and $Q(\phi, \nu)$ using $\varepsilon$-$greedy$.}
\STATE{Take action $\nu_k$, observe the overall service profit $R^s_k$} over $T$ time slots in $N_k$ under task scheduling policy $\pi^f$ in the fast timescale.
\STATE{Observe system state $\phi_{k+1}$ at the beginning of $N_{k+1}$.}
\STATE{Update action-value $Q(\phi_{k}, \nu_k)$ according to Eq. \eqref{eqn:q_learning}}
\STATE{$k \gets k+1$ }
\UNTIL{$k<M$}
\end{algorithmic}
\end{algorithm}

\section{System Implementation and Performance Evaluation} \label{sec_6}



\subsection{System Implementation and Experiment Settings}
We implement our cloud video transcoding system \emph{Morph} in Python. The source code of our project is released in Github, which can be found in \cite{akilos}.
We run the transcoding system in a cloud environment built with Docker.
We use the ffmpeg for the video transcoding operation.
The price for renting a VM instance is \$0.252 per hour.
Each of the VM instances in our home-built cloud has 4 CPU cores, the CPU frequency is 2.10 GHz, and the memory size is 2GB. The VM instances are connected with Gigabit Ethernet, and the data transmission speed can achieve 1000 Mbit/s among the VM instances.
Our system provides three levels of services, denoted as Level \uppercase\expandafter{\romannumeral1}, Level \uppercase\expandafter{\romannumeral2}, and Level \uppercase\expandafter{\romannumeral3},  respectively. The initial marginal price $R_i$ is \$0.018 per minute for service level \uppercase\expandafter{\romannumeral1}, \$0.012 per minute for service level \uppercase\expandafter{\romannumeral2}, and \$0.006 per minute for service level \uppercase\expandafter{\romannumeral3}. The price discounting factor $\alpha$ is 0.999 per second.
%
The consumed computing time for each video block is 180 seconds.

\subsection{Transcoding Time Estimation Accuracy}

\begin{figure}
\centering
\subfigure[\small{Error histogram of neural network method.}]{\includegraphics[width=1.725in]{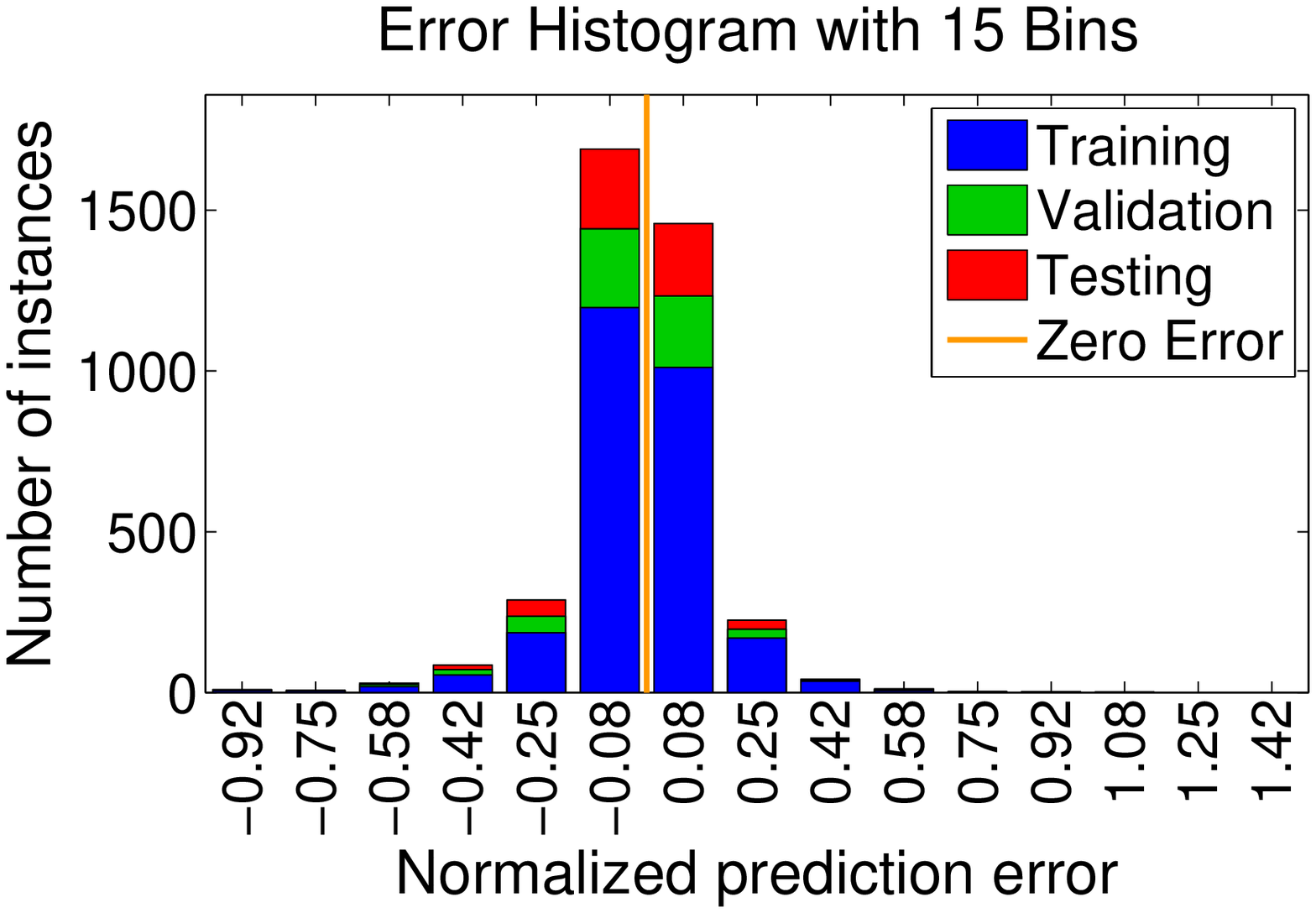}\label{nn_error}}
\subfigure[\small{Error histogram of linear approximation method.}]{\includegraphics[width=1.725in]{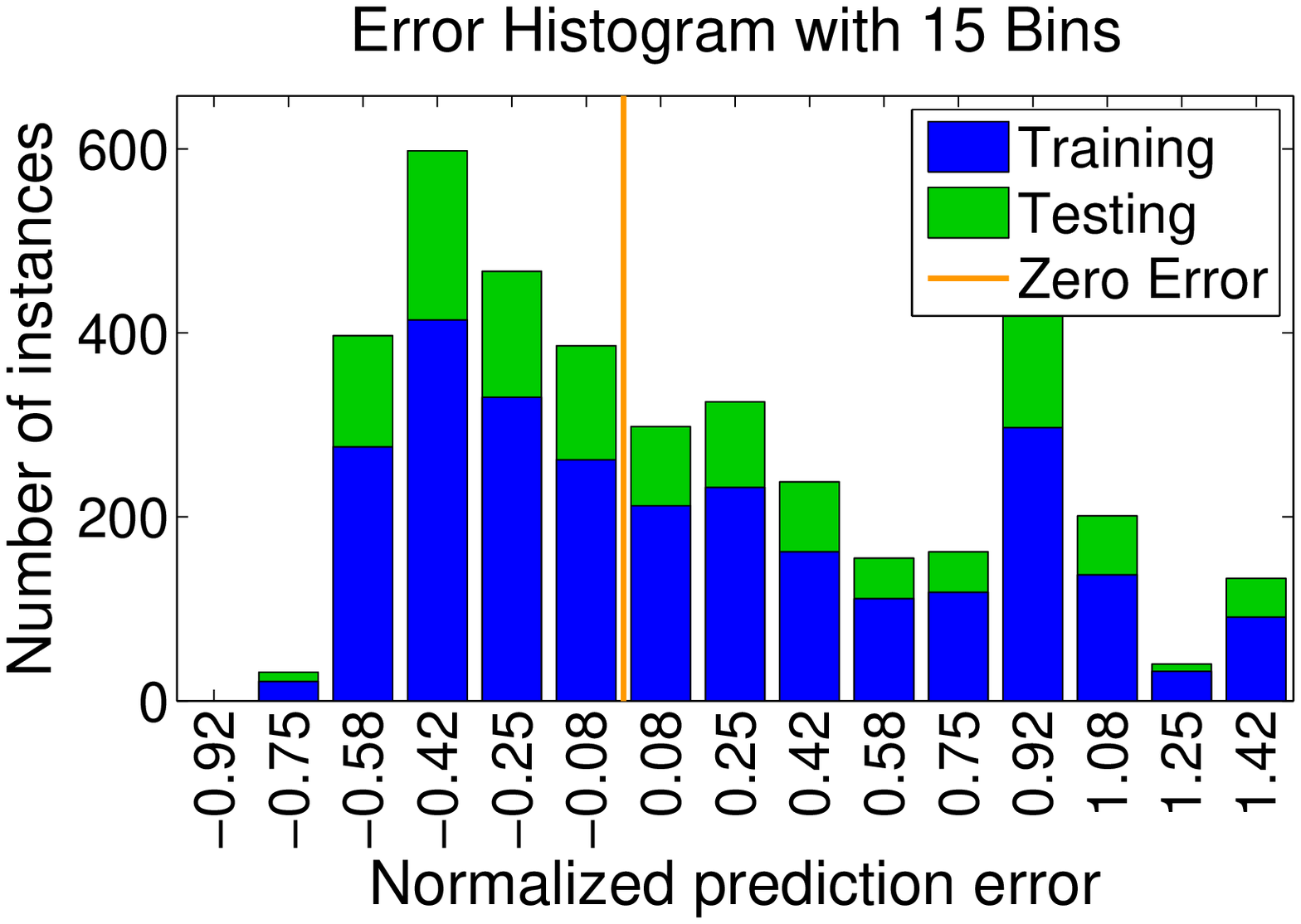}\label{linear_approximation_hist}}
\caption{Comparison between the transcoding time prediction methods.}
\end{figure}

We measure the time for transcoding 2020 video files of different original bitrates and resolutions into three target resolutions, namely, 854x480, 640x360, 426x240.
We obtain 3850 instances of the measured transcoding time.
We select 75\% of the data for training the neural network offline, 15\% of the data for model validation, and the other 15\% of the data for testing. The hidden layer of the neural network consists of 20 neurons. The inputs of the neural network include the video bitrate, resolution, frame rate, duration of the original video file, and the target video resolution. We multiply the width and height of the video resolution and use the product as one input.
We compare the neural network method for transcoding time estimation with the linear approximation method which estimates transcoding time as a linear function of the video duration.
We normalize the prediction error of the test using the following equation for comparison
\begin{equation}
Normalized \ Error = \frac{Predicted \ Time - Real \ Time}{Real \ Time}. \nonumber
\end{equation}
The transcoding time is measured in seconds in our experiments.
The error histogram of the neural network method for transcoding time prediction is illustrated in Fig. \ref{nn_error}.
The normalized prediction error of most of the testing instances are within the range from -0.08 to 0.08.
%
%
The error histogram of the linear approximation method is illustrated in Fig. \ref{linear_approximation_hist}, the normalized prediction error of the testing instances ranges from -0.58 to 1.42.
From the comparisons, we can observe that the neural network method can predict the transcoding time much more precisely.

\subsection{Service Profit under Real Trace Data}

%


We measure the service profit under a real trace data. The trace data captures the video requests to a CDN node. We extract the user requests in the trace data as the transcoding requests for our service.
We divide the time of each day into 24 hours and average the request number during one hour of the days as the average task arrival rate in the system state for this hour of a day.
%
We scale down the average request rate in the real trace to the range of 0.1-0.7 request per minute.
%


\begin{figure}
\centering
\subfigure[\small{The cumulative profit for different methods over 24 hours.}]{\includegraphics[width=1.725in]{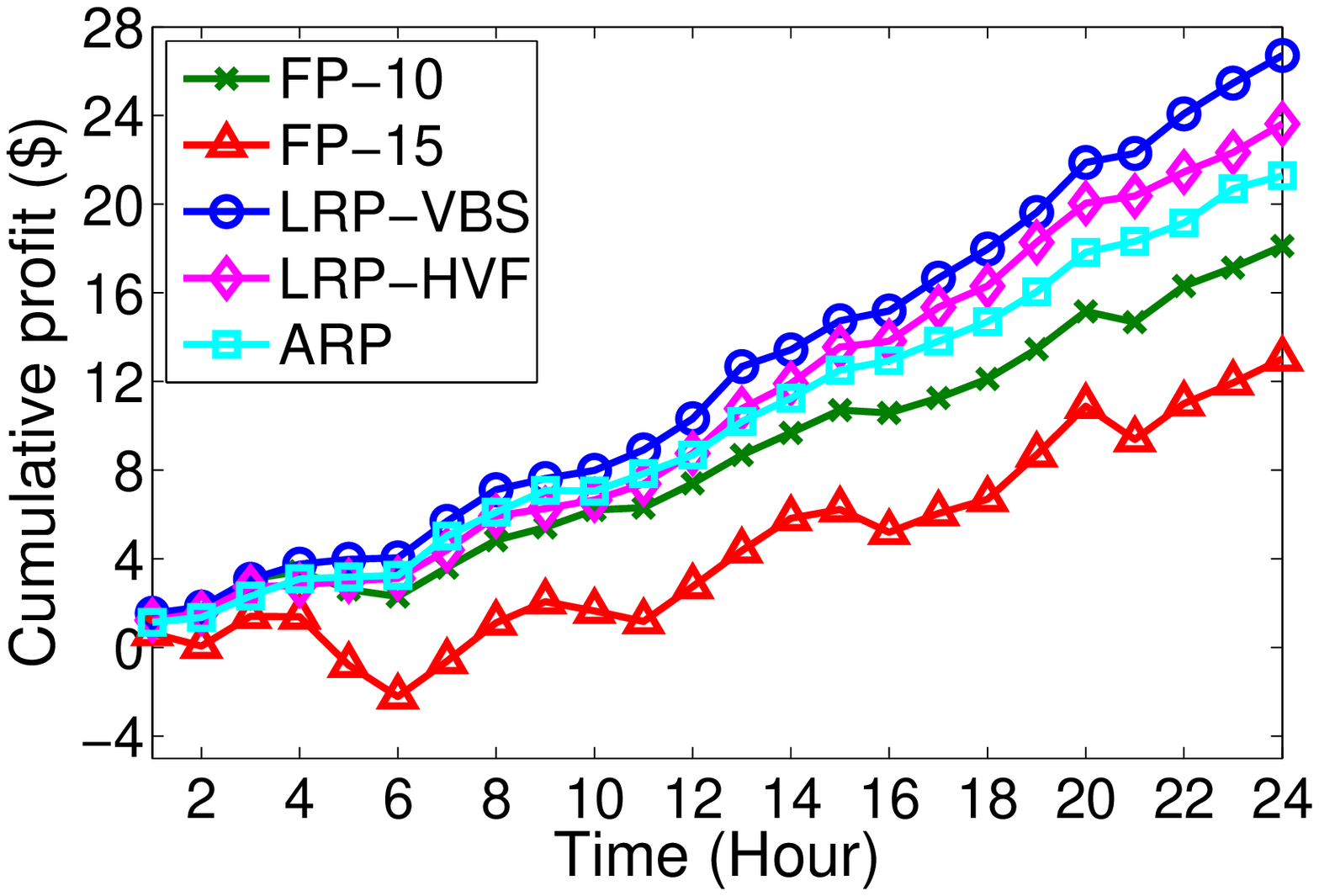}\label{service_profit}}
\subfigure[\small{The number of provisioned VM instances in each hour.}]{\includegraphics[width=1.725in]{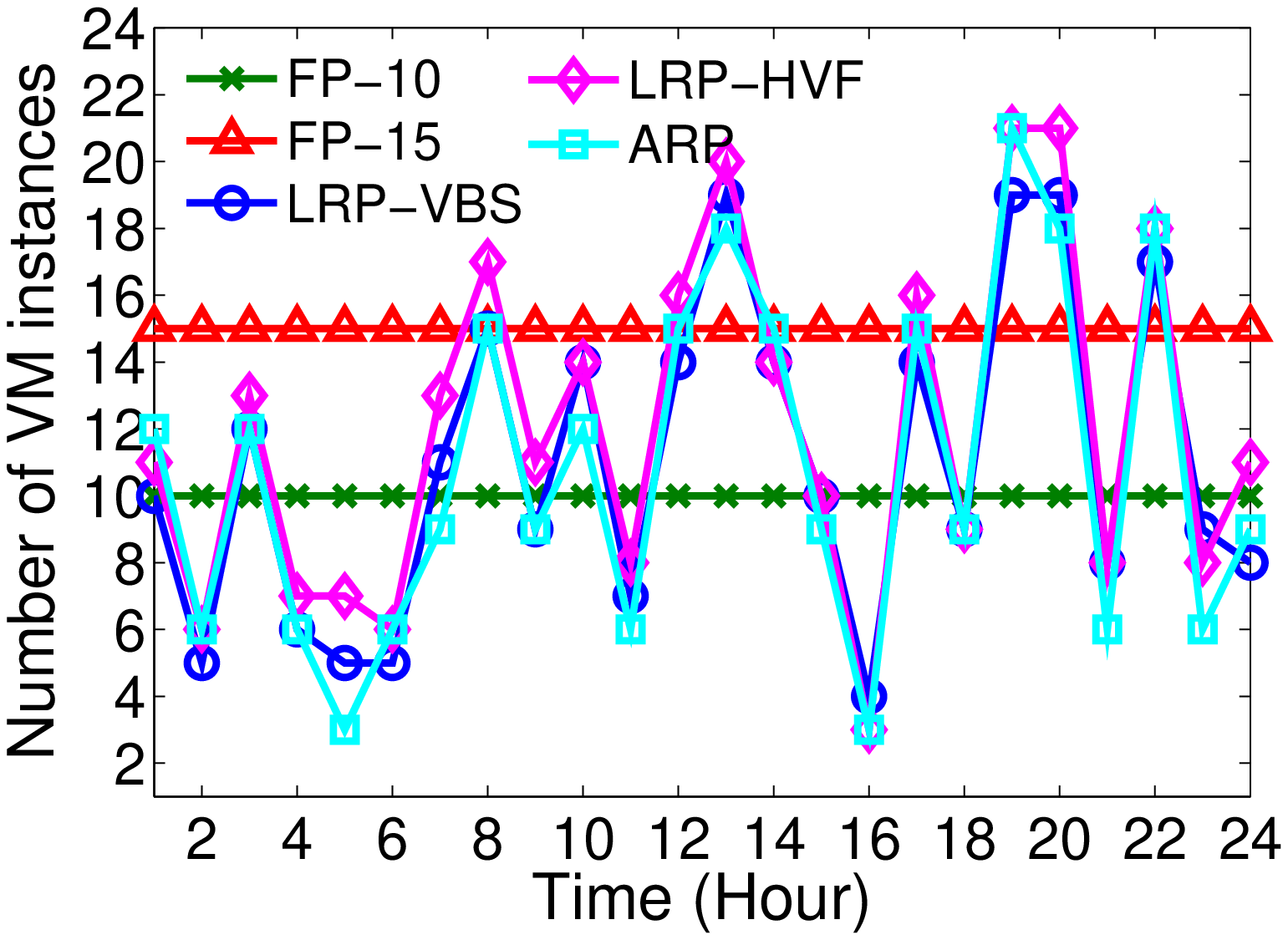}\label{Epoch_instances}}
\caption{System performance under the real trace.}
\end{figure}

We measure the service profit in the real environment over 24 hours under different resource provisioning policies and task scheduling policies.
%
%
We refer to the value-based task scheduling policy as VBS.
We compare the service profit under our proposed Learning-based Resource Provisioning policy and Value-based Task Scheduling (LRP-VBS) policy with the following methods:
1) The Fixed Policy (FP) which runs a fixed number of VM instances under the task scheduling policy of VBS. We select two representative numbers of VM instances for the FP methods to illustrate, i.e., 10 VM instances and 15 VM instances.
2) The Arrival Rate based Policy (ARP), i.e.,
the number of provisioned VM instances is proportional to task arrival rate. This method estimates the number of VM instances to satisfy the computing resource requirements of the transcoding tasks in each period based on the task arrival rate.
In our test, we set the number of provisioned VM instances as 30 times of the task arrival rate per minute.
3) The learning-based resource provisioning policy combined with other task scheduling policy.
We select the learning-based resource provisioning policy combined with the task scheduling policy of Highest Value First (LRP-HVF). The HVF policy always selects the transcoding task which has the current highest valuation to perform. The current valuation of the tasks are calculated by Eq. \eqref{valuation_function}.

We illustrate the cumulative profit under different methods over 24 hours in Fig. \ref{service_profit}.
We can observe that the cumulative service profit under our proposed LRP-VBS is larger than the baseline methods.
As demonstrated in Fig. \ref{Epoch_instances}, the LRP method can effectively adjust the number of provisioned VM instances in the slow timescale given the task scheduling policy HVF or VBS in the fast timescale.
The task scheduling policy in the fast timescale will also affect the resource provisioning operation in the slow timescale.
As such, LRP-VBS can gain a higher profit than the LRP-HVF since that the VBS can outperform the HVF for task scheduling in the fast timescale.
In constrast, the FP method may waste much computing resource when the system workload is low and deteriorate the system performance when the system workload exceeds the processing capacity of the current number of VM instances.
The ARP method can dynamically adjust the number of provisioned VM instances according to the task arrival rate and system workload, however, it is hard for this method to model the decreasing of the revenue for the processing delay.
Moreover, the service revenue for the transcoding tasks is also affected by the task scheduling policy, but the ARP method does not take the task scheduling scheme into consideration.
Therefore, the ARP method cannot effectively determine the optimal policy for maximizing the service profit.
The LRP is more suitable for such hard-to-model problem, and it can work effectively in the system dynamic by learning the optimal policy during the training stage.

\section{Conclusions} \label{sec_7}
We consider the problem of how to provision the computing resource for transcoding the video contents while maximizing the service profit for the transcoding service provider.
We design a practical transcoding system by leveraging the cloud computing infrastructure.
We jointly consider the task scheduling and resource provisioning problem in the two timescales and
formulate the service profit maximization problem as a two-timescale MDP.
We derive the approximate solutions for task scheduling and resource provisioning.
Based on our proposed methods,
we implement the system and conduct extensive experiments to evaluate the system performance in a real environment.
The experiment results show our method can work effectively for the task scheduling and resource provisioning in the cloud video transcoding system.


\appendices
\section{Proof of Proposition \ref{prop_1}} \label{append_1}
We denote the remaining time to complete transcoding the current video block on the VM instance $i$ as $y_i$.
We assume that at $t_0$, the transcoding worker $1$ becomes idle and requests the first video block in the queue, and therefore $y_1=F$.
The waiting time for the next video block to be requested is
\begin{eqnarray}
Y = \min\{y_2, y_3, ..., y_{m_k}\}, \nonumber
\end{eqnarray}
where $y_2, y_3,...,y_{m_k}$ are unknown and randomly distributed in $[0,\ F]$ and $0 \le Y \le F$.
The transcoding progresses on the VM instances are independent and the CDF of $Y$ is
\begin{align}
F_Y(t)& = P(0 \le Y \le t) \nonumber \\
      & = 1 - P(y_2>t)P(y_3>t)...P(y_{m_k}>t). \nonumber
\end{align}
Hence, the expected waiting time for the next block to be requested can be calculated as $E\{Y\} = \frac{F}{{m_k}}$.
We can deduce that the total waiting time for the $g_i$-th video block to be requested by a transcoding worker is
$\frac{F}{{m_k}}(g_i - 1)$.
As such, the estimated completion time of the $g_i$-th video block is
\begin{eqnarray}
E\{f_i\} = t_0 + \frac{F}{{m_k}}(g_i - 1) + F.
\end{eqnarray}



%

\bibliographystyle{IEEEtran}
\bibliography{transcoding}

\begin{thebibliography}{1}
\providecommand{\url}[1]{#1}
\csname url@samestyle\endcsname
\providecommand{\newblock}{\relax}
\providecommand{\bibinfo}[2]{#2}
\providecommand{\BIBentrySTDinterwordspacing}{\spaceskip=0pt\relax}
\providecommand{\BIBentryALTinterwordstretchfactor}{4}
\providecommand{\BIBentryALTinterwordspacing}{\spaceskip=\fontdimen2\font plus
\BIBentryALTinterwordstretchfactor\fontdimen3\font minus
  \fontdimen4\font\relax}
\providecommand{\BIBforeignlanguage}[2]{{%
\expandafter\ifx\csname l@#1\endcsname\relax
\typeout{** WARNING: IEEEtran.bst: No hyphenation pattern has been}%
\typeout{** loaded for the language `#1'. Using the pattern for}%
\typeout{** the default language instead.}%
\else
\language=\csname l@#1\endcsname
\fi
#2}}
\providecommand{\BIBdecl}{\relax}
\BIBdecl

\bibitem{fayazbakhsh2013less}
S.~K. Fayazbakhsh, Y.~Lin, A.~Tootoonchian, A.~Ghodsi, T.~Koponen, B.~Maggs,
  K.~Ng, V.~Sekar, and S.~Shenker, ``Less pain, most of the gain: Incrementally
  deployable icn,'' \emph{ACM SIGCOMM Computer Communication Review}, vol.~43,
  no.~4, pp. 147--158, 2013.

\bibitem{wen2014cloud}
Y.~Wen, X.~Zhu, J.~Rodrigues, and C.~Chen, ``Cloud mobile media: Reflections
  and outlook,'' \emph{Multimedia, IEEE Transactions on}, vol.~16, no.~4, pp.
  885--902, 2014.

\bibitem{tschudin2013named}
C.~Tschudin and M.~Sifalakis, ``Named functions for media delivery
  orchestration,'' in \emph{Packet Video Workshop (PV), 2013 20th
  International}.\hskip 1em plus 0.5em minus 0.4em\relax IEEE, 2013, pp. 1--8.

\bibitem{Song:2015}
\BIBentryALTinterwordspacing
M.~Song, Y.~Lee, and J.~Park, ``Scheduling a video transcoding server to save
  energy,'' \emph{ACM Trans. Multimedia Comput. Commun. Appl.}, vol.~11,
  no.~2s, pp. 45:1--45:23, Feb. 2015. [Online]. Available:
  \url{http://doi.acm.org/10.1145/2700282}
\BIBentrySTDinterwordspacing

\bibitem{ma2014dynamic}
H.~Ma, B.~Seo, and R.~Zimmermann, ``Dynamic scheduling on video transcoding for
  mpeg dash in the cloud environment,'' in \emph{Proceedings of the 5th ACM
  Multimedia Systems Conference}.\hskip 1em plus 0.5em minus 0.4em\relax ACM,
  2014.

\bibitem{timmerercloud}
C.~Timmerer, D.~Weinberger, M.~Smole, R.~Grandl, C.~M{\"u}ller, and S.~Lederer,
  ``Cloud-based transcoding and adaptive video streaming-as-a-service,''
  \emph{E-LETTER}.

\bibitem{chang2003multitime}
H.~S. Chang, P.~J. Fard, S.~Marcus, M.~Shayman \emph{et~al.}, ``Multitime scale
  markov decision processes,'' \emph{Automatic Control, IEEE Transactions on},
  vol.~48, no.~6, pp. 976--987, 2003.

\bibitem{akilos}
``{Morph: Cloud Video Transcoding},'' \url{https://github.com/cap-ntu/Morph},
  [Online; accessed Dec-2015].

\end{thebibliography}

\end{document}